\documentstyle[aps,prbbib]{revtex}
\begin{document}
\newcommand{\be}{\begin{equation}}
\newcommand{\ee}{\end{equation}}
\newcommand{\p}{^{\prime}}
\newcommand{\f}{\displaystyle\frac}
\title{With the relativistic velocity addition law through special relativity}

\author{B.Rothenstein{\footnote{Corresponding author:
bernhard\_rothenstein@yahoo.com}}, I.Zaharie\\ {\small Physics
Department, "Politehnica" University Timi\c soara,}\\ {\small Pia\c ta
Regina Maria, nr.1, 1900 Timi\c soara, Romania}}
\date{}
\maketitle
\begin{abstract}

It is shown that if we can define a physical quantity with proper character  in a given inertial
reference frame (kinematic, dynamic, electromagnetic in its nature) which transforms when detected
from a reference frame relative to which it moves with velocity $u_x$ as
$F=\f{F^o}{\sqrt{1-\f{u_x^2}{c^2}}}$ then we can derive for it transformation equations following
one and the same procedure, which involves the addition law of relativistic velocities which can be derived
without using the Lorentz transformations. The transformation equation derived that way, generates
the physical quantities $u_xF$ and $u_x'F'$, for which physicists invent names  reflecting theirs physical meaning.
\end{abstract}

\section{Introduction }

Each  chapter of physics operates with fundamental physical quantities. Kinematics is involved
with length and time, dynamics add to them, by tradition, the concept of mass, whereas
electromagnetism introduces the concept of electric charge. Combining fundamental physical
quantities we obtain new ones,  useful in describing physical effects.

All chapters of physics have in common the relativistic postulate, due to Galileo and Einstein,
according to which all the true laws of physics are the same  for all inertial observers, all of
them measuring the value $c$ for the velocity of light.
When measured from different reference frames some physical quantities have the same magnitude
(relativistic invariants). If a physical quantity which characterizes an object is measured by
an observer relative to whom the  object is in a state of rest, we say that the observer
measures its proper value.

An equation which relates physical quantities measured from different reference frames,
represents a transformation equation. A transformation equation enable us to derive formulas
which account for different relativistic effects. It is considered  that the use of the
transformation equations reduces the transparency $^{1,2}$.

Relativistic equations describing time dilation$^{1,3,4,5}$, length contraction and addition of
relativistic velocities$^{1,6,7}$ can be derived without using transformation equations, all
having as a starting point the time dilation formula. The method works when in one of the involved
reference frames a proper physical quantity is measured.

The purpose of our paper is to show that the transformation equations for kinematic, dynamic and
electromagnetic physical quantities can be derived using the same strategy, involving the addition
law of relativistic velocities as a single relativistic ingredient, which at its turn is a direct
consequence of the time dilation effect$^7$.

Three inertial reference frames are involved $S(XOY)$, $S'(X'O'Y')$ and $S^o(X^oO^0Y^0)$.
$S^o$ moves with velocity $u_x$ relative to $S$, with velocity $u_x'$ relative to $S'$, $S'$ moving
at its turn with velocity $v$ relative to $S$. The axes of the reference frames are parallel to
each other, the $OX$, $O'X'$ and $O^oX^o$ axes are overlapped and all the velocities show in
theirs  positive directions. The three reference frames have the same space-time origins.
In order to keep the problem as simple as possible, we will consider only events which take place
on the overlapped axes. The velocities mentioned above are related by
\be
u_x = \f{u_x' + v}{1+\f{u_x'v}{c^2}}
\ee
which solved for $u_x'$ leads to
\be
u_x' = \f{u_x - v}{1-\f{u_xv}{c^2}}.
\ee
Reference frame $S^o(X^oO^0Y^0)$ is a special one. In that frame nothing and nobody moves and
only events which take place at it origin $O^o$ are taken into account. The result is  that there,
only physical quantities which characterize a stationary body can be detected, like the reading
of clock $C_o^o$ located at $O^o(t^o)$, the mass $m^o$ of a particle located at $O^o$, the force
$F^o$ acting on the particle mentioned above and the electric field created by a charged capacitor.
Physical quantities like different from zero space coordinates, momentum, mechanical power and
magnetic field are meaningless in that frame.
All the physical quantities measured in $S^o$ are by definition
proper physical quantities: proper time reading (date) $t^o$, proper mass $m^o$, proper force
$F^o$ and proper electric field $E^o$.

\section{Relativistic kinematics }

In each of the mentioned reference frames, the clocks located  at the different points of the
overlapped axes are synchronized following a procedure proposed by Einstein.
The proposed scenario makes that at a given point of the overlapped axes clock $C_o^o$ located
at $O^o$ meets a clock $C(x,y=0)$ of the $S$ frame and a clock $C'(x',y'=0)$ of the $S'$ frame.
Let $t^o$, $t$ and $t'$ be the readings of the three clocks respectively. In
accordance with the time dilation formula the three readings are related by
\be
t = \f{t^o}{\sqrt{1-\f{u_x^2}{c^2}}}
\ee
and by
\be
t' = \f{t^o}{\sqrt{1-\f{u_x'^2}{c^2}}}.
\ee
Eliminating $t^o$ we obtain that $t$ and $t'$ are related by
\be
t = t' \f{\sqrt{1-\f{u_x'^2}{c^2}}}{\sqrt{1-\f{u_x^2}{c^2}}} = t'\f{1+\f{u_x'v}{c^2}}{\sqrt{1-\f{v^2}{c^2}}}=
\f{t'+\f{v(u_x't')}{c^2}}{\sqrt{1-\f{v^2}{c^2}}}
\ee
and by
\be
t' = t \f{\sqrt{1-\f{u_x^2}{c^2}}}{\sqrt{1-\f{u_x'^2}{c^2}}} = t\f{1-\f{u_xv}{c^2}}{\sqrt{1-\f{v^2}{c^2}}}=
\f{t-\f{v(u_xt)}{c^2}}{\sqrt{1-\f{v^2}{c^2}}}.
\ee
We detect in the right hand side of equations (5) and (6)  the presence of the terms ($u_xt$)
and ($u_x't'$) respectively. We have to find out names for them. Taking into account that
clock $C_o^o$ is located at the common origin of time at the common origins of the reference
frames, $x=u_x t$ can represent the space coordinate of clock $C_o^o$ in the $S$ frame when the
synchronized clocks of that frame read $t$ or the distance travelled by $C_o^o$ during the time
interval ($t-0$), measured of course in $S$. For the same reasons, $x'=u_x' t$ can represent
the distance travelled by $C_o^o$ in the time interval ($t'-0$), measured  in $S'$ or the
space coordinate of clock $C_o^o$ in the $S'$ frame when the synchronized clocks read there $t'$.
With this new notations equations (5) and (6) become
\be
t=\gamma_v \left(t'+\f{v x'}{c^2}\right) ; \hspace{2cm}\gamma_v = \f{1}{\sqrt{1-\f{v^2}{c^2}}}
\ee
\be
t'=\gamma_v \left(t-\f{v x}{c^2}\right).
\ee
The space coordinates $x$ and $x'$ transform as
\be
x = u_x t = \gamma_v t'(u_x' + v) =  \gamma_v (x' + v t')
\ee
and
\be
x' = u_x' t' = \gamma_v t(u_x - v) =  \gamma_v (x - v t).
\ee
We have derived so far the Lorenz transformations for the events $E(x=u_x t, y=0, t)$ associated
with the fact that clock  $C_o^o$ meets clock $C(x=u_x t, y=0)$ and $E'(x'=u_x' t', y'=0, t')$
associated with the fact that clock $C_o^o$ meets clock $C'(x'=u_x' t', y'=0)$. Because the two events
take place at the same point in space, relativists say that they represent the same event.

\section{Relativistic dynamics }

Consider that a particle is located at rest at the origin $O^o$ of $S^o$. Observers in that frame
measure its rest mass $m^o$. Using the addition law of relativistic velocities it was
shown$^{13}$ that as measured from $S$ the mass of the particle is
\be
m = \f{m_0}{\sqrt{1-\f{u_x^2}{c^2}}}
\ee
whereas as measured from $S'$ it is
\be
m' = \f{m_0}{\sqrt{1-\f{u_x'^2}{c^2}}},
\ee
resulting that $m$ and $m'$ are related by
\be
m = m'\f{\sqrt{1-\f{u_x'^2}{c^2}}}{\sqrt{1-\f{u_x^2}{c^2}}} = m' \f{1+\f{vu_x'}{c^2}}{\sqrt{1-\f{v^2}{c^2}}}=
\gamma_v \left(m' + \f{v(m'u_x')}{c^2}\right)
\ee
and by
\be
m' =  m\f{1-\f{vu_x}{c^2}}{\sqrt{1-\f{v^2}{c^2}}} =
\gamma_v \left(m - \f{v(mu_x)}{c^2}\right).
\ee
It is considered that $m$ and $m'$ represent the relativistic mass of the same particle as measured
from $S$ and $S'$ respectively. We detect in the right hand side of equations (13) and (14) the
presence of the terms $mu_x$ and $m'u_x'$ which have the physical dimensions of momentum.
Physicists baptize physical quantities by a single name and label them by a single notation.
So they introduce the notations
\be
p_x = m u_x = \f{m^o u_x}{\sqrt{1-\f{u_x^2}{c^2}}}
\ee
\be
p_x' = m' u_x' = \f{m^o u_x'}{\sqrt{1-\f{u_x'^2}{c^2}}},
\ee
$p_x$ and $p_x'$ representing the relativistic momentum of the particle with rest mass $m^o$ as
measured from $S$ and $S'$ respectively.

The relativistic momentum transforms as
\be
p_x = m u_x = \gamma_v m' (u_x' + v) = \gamma_v (p_x' + v m')
\ee
\be
p_x' = m' u_x' = \gamma_v m (u_x - v) = \gamma_v (p_x - v m).
\ee
The physical quantity $E=mc^2$ ($E'=m'c^2$) has the physical dimensions of energy. Due to the
invariance of $c$, it transform as mass does
\be
E = E' \gamma_v \left(1+\f{vu_x'}{c^2}\right) = \gamma_v (E' + v p_x')
\ee
and
\be
E' = E \gamma_v \left(1-\f{vu_x}{c^2}\right) = \gamma_v (E - v p_x).
\ee
Expressed as a function of energy, equations (19) and (20) become
\be
p_x = \gamma_v \left(p_x' + \f{v E'}{c^2}\right)
\ee
and
\be
p_x' = \gamma_v \left(p_x - \f{v E}{c^2}\right).
\ee
Equations (19) and (22) put the bane on the concept of relativistic mass as many
physicists require$^8$.

A remarkable property of the relativistic energy consists in the fact that a particle at rest is
characterized by a rest energy
\be
E^o = m^o c^2.
\ee
Because the single supplementary energy a free particle moving with constant velocity can possess
is its kinetic energy $E_k$, we have for it
\be
E_k = E - E_o = E_o(\gamma_v - 1) = m^o c^2 (\gamma_v - 1).
\ee
It is clear that $p_x$ ($p_x'$), $E$ ($E'$) express the momentum and the energy of the same particle
as measured from $S$ and $S'$.

The traditional way to derive the equations presented above is to consider a collision between
two particles from two inertial reference frames and imposing that momentum and energy conserve$^9$.

The truth of equation (11) is confirmed by the experiments performed Bucherer$^{10}$ with electrons
moving in a magnetic field, obtaining excellent agreement working with it. This was and continues to be
a great proof in favour of the special theory of relativity in all its fields, all being the
consequence of Einstein's postulate of relativity. It is considered$^{10}$ that to this date no
other explanation of Bucherer's experiment then that supplied by special relativity, is known.

\section{Relativistic electrodynamics }

We consider now a capacitor $C_o$ the plates of which are parallel to the $X^oO^oZ^o$ plane, at rest
in $S^o$ and charged with electrical charges $Q$ and $-Q$ respectively. It creates an electric
field $E_y^o$, oriented in the positive direction of the $O^oY^o$ axis given by
\be
E_y^o = \f{Q}{\varepsilon_0 \varepsilon_r A^o},
\ee
$A^o$ representing the surface of one of the plates $\varepsilon_0$ and $ \varepsilon_r$ representing
invariant electrical constants. The electrical  neutrality of atoms convinces us that the electric
charge is a relativistic invariant. When detected from $S$, the sides of the plate parallel with the
overlapped axes undergo a Lorentz contraction and so the electric field measured from $S$ is
\be
E_y = \f{Q}{\varepsilon_0 \varepsilon_r A^o \sqrt{1-\f{u_x^2}{c^2}}}
\ee
whereas in $S'$ it is
\be
E_y' = \f{Q}{\varepsilon_0 \varepsilon_r A^o \sqrt{1-\f{u_x'^2}{c^2}}}.
\ee
The same procedure, as in the previous cases leads to the following relationship between $E_y$
and $E_y'$
\be
E_y = E_y' \gamma_v\left(1+\f{vu_x'}{c^2}\right) = \gamma_v \left(E_y' + \f{v(u_x'E_y')}{c^2} \right)
\ee
\be
E_y' = E_y \gamma_v\left(1-\f{vu_x}{c^2}\right) = \gamma_v \left(E_y - \f{v(u_xE_y)}{c^2} \right).
\ee
We detect in the right hand side of equations (28) and (29) the presence of the terms ($u_xE_y$)
and ($u_x'E_y'$) and label them $B_z$ and $B_z'$ respectively, representing the $OZ$ ($OZ$)
components of a vector $\vec B (0, 0, B)$ in $S$ and  $\vec B' (0, 0, B')$ in $S'$, baptized
magnetic field. Its direction is chosen in such a way that the vector products
$\vec u_x \times \vec E_y$ and $\vec u_x' \times \vec E_y'$ show in the same direction as
$E_y$ and $E_y'$ do, respectively. It transform as
\be
B_z = \gamma_v E_y'(u_x'+ v) = \gamma_v (B_z' + v E_y')
\ee
\be
B_z' = \gamma_v E_y(u_x- v) = \gamma_v (B_z - v E_y).
\ee
Expressed as a function of the magnetic field, equations (28) and (29) become
\be
E_y = \gamma_v \left(E_y' + \f{v B_z'}{c^2} \right)
\ee
\be
E_y' = \gamma_v \left(E_y - \f{v B_z}{c^2} \right).
\ee
With the plates of the capacitor parallel to the $Y^oO^oZ^o$, it creates an electric field parallel
to the overlapped axes. In that case the sides of the plates are not subjected to the Lorentz
contraction and so
\be
E_x^0 = E_x = E_x'.
\ee
The transformation  equations for the components $E_z$ ($E_z'$) and $B_y$ ($B_y'$) are obtained
starting with the capacitor the plates of which are parallel to the $XOY$ plane producing an
electric field orientated along the $OZ$ axis.

The traditional  way to derive the equations presented above is to use the transformation
properties of force$^{11,12}$.

\section{Conclusions }

We have shown that if a proper physical quantity  $F^o$ (proper time, proper mass, proper electric
field) can be defined in a given reference frame $S^o$ which moves  with velocity $u_x$ relative
to $S$ and with velocity $u_x'$ relative to $S'$ and transforms as
\be
F = \f{F^o}{\sqrt{1-\f{u_x^2}{c^2}}}
\ee
\be
F' = \f{F^o}{\sqrt{1-\f{u_x'^2}{c^2}}}
\ee
respectively, then a transformation equation relating $F$ and $F'$ can be derived
using the addition law of velocities as a single relativistic ingredient. The transformation
generate a new physical quantity ($uF$), ($u'F'$) in $S$ and in $S'$ respectively for which
physicists find adequate names and the transformation of which involves the transformation
of $F$ and $F'$ and the addition law of velocities.

As compared with the traditional derivations which are very laborious, our derivation is heuristic,
being based on facts known from classical physics, uses simple algebra being time saving having a
good  relationship between time invested and understanding achieved.

We consider that it convinces the reluctants  about the truth of special relativity. It
does not charge the memory of the teacher or of the learner, who in the traditional approaches,
have to learn in each chapter of physics, a new scenario.


\end{document}